# Direct Visualization of Temperature-Induced Phase Separation of Completely Miscible Au-Pd Alloy by In-Situ TEM

*Abhijit Roy\*, Simon Hettler, and Raul Arenal\**

A. Roy, S. Hettler, R. Arenal

Laboratorio de Microscopías Avanzadas (LMA), Universidad de Zaragoza, Zaragoza, Spain

E-mail: aroy@unizar.es ;    arenal@unizar.es

A. Roy, S. Hettler, R. Arenal

Instituto de Nanociencia y Materiales de Aragón (INMA), CSIC-Universidad de Zaragoza, Zaragoza 50009, Spain

R. Arenal

ARAID Foundation, Zaragoza 50018, Spain

In-situ transmission electron microscopy (TEM) studies reveal key insights into the structural and chemical evolution of nanoparticles (NPs) under external stimuli like heating and biasing, which is critical for evaluating their suitability in chemical reactions and their tendency towards forming novel NP systems. In this study, starting from a core@shell Au nanotriangle (AuNT)@Pd nanostructure, the formation of a phase-separated bi-metallic Au-Pd NP system at high temperature is reported, despite the fact that Au and Pd are miscible in the entire composition and temperature range. In-situ TEM heating of bare AuNT@Pd core@shell structures up to 1000°C has been performed. Between 400°C and 800°C, an initial alloy formation has been observed. It is also noted that higher initial loading of Pd increases the melting temperature of the bi-metallic system. However, the most important observation is the separation of the nanostructure into Au and Pd phases at temperatures higher than 850° C for high Pd doping. The extent of Pd separation depends on the amount of initial Pd loading. A Janus Au-Pd nanostructure is formed at the end of the thermal treatments at 1000ºC. The phase-separated NP is observed to be highly stable and could be clearly beneficial for various applications, particularly in catalytic processes.

1. Introduction

Bi-metallic nanoparticles (NPs) have garnered significant attention for their superior efficiency in heterogeneous catalysis and their potential for a transition towards green energy production, as well as for their uses in optical and magnetic applications.[1,2] Bi-metallic systems in different configurations showed superior stability, selectivity and activity compared to any of their individual metallic counterparts.[1–4] In this aspect, Au@Pd core@shell NPs and, more recently, Au-Pd nano-alloys and supported Au/Pd bi-metallic nanocatalysts offered superior performance in the electrocatalytic production of hydrogen peroxide ($H_2O_2$), ethanol oxidation reactions, formic acid dehydrogenation,[5] alcohol to aldehyde conversion,[6] CO oxidation[7] and acetylene trimerization.[8] Moreover, a comparison with bare Pd NPs revealed a 1.5- to 3-fold increase in catalytic activity.[9–12] This superior catalytic activity is attributed to a combination of ensemble and ligand effects.[13] Furthermore, Au@Pd core@shell NPs showed superior photocatalytic properties for the generation of photocatalytic $H_2$ using solar energy as the activation force.[14–18] In these core@shell photocatalytic applications, the inner Au core acts as the energy (or light) absorber, which generates hot electrons ($e^-$) and holes ($h^+$) in the Pd shell. These hot electrons and holes are available for the reduction reaction.

As mentioned above, the alloyed bi-metallic configuration is mostly used for electrochemical reduction reactions. It is worth mentioning that Au NPs are less efficient for oxygen reduction but highly efficient for conversion of alcohols to aldehyde, which is completely opposite to the catalytic activity of Pd.[19] These limitations have been addressed by creating alloyed systems of Au and Pd. However, it has been observed that the composition and structure of Au@Pd nanoalloys and core@shell structures vary considerably during heating and under the reaction conditions.[20–23] Also, the lower electrochemical potential of these alloys, in contrast to the individual metals, decreases their efficiency towards alcohol-to-aldehyde conversion reaction.[24,25] Thus, it should be highly beneficial to obtain a bi-metallic Au-Pd system where both phases are separated from each other in a single NP. In such systems, the oxidation and reduction reactions will take place at different locations of the same NP, leading to superior catalytic activity and stability. Such systems are unique in the sense that they retain the different properties of the individual constituents but additionally provide physical, chemical and optical interactions between them due to the presence of a common interface.[26] Previously, phase-separated bi-metallic systems consisting of Au-Pt, Au-Ni, Au-Fe, and Ga-In were obtained very efficiently. These systems showed superior catalytic and improved optical properties compared to their individual counterparts.[27–29] For instance, it has been shown that Pt–Ni octahedral nanocrystals in the oxygen reduction reaction (ORR) can surpass those of pure Pt counterparts by nearly 20 times.[30] However, obtaining bi-metallic NPs based on the phase-separation method depends on the fact that the lattice mismatch between two constituent metals must be large (>5%, ex. Au-Ni, Au-Cu, Ga-In). Metals with lower lattice mismatch tend to create an alloyed system due to complete miscibility for the entire composition range at all the temperatures according to their bulk phase diagram (i.e. Au-Pd).[31] Though researchers theoretically predicted a miscibility gap of an Au-Pd bi-metallic system, the conversion of an Au-Pd nanoalloy to a phase-separated Au/Pd bi-metallic system with annealing could not be obtained till now.[32–34] Au/Pd bi-metallic Janus nanostructures were previously reported using a wet chemical method at room temperature.[35–38] In this approach, the surfactants and reducing agents were observed to play the most dominant role behind the asymmetric growth of the NPs. However, these NPs are obtained by controlling the reaction kinetics to reach a local minimum of the Gibbs free energy. Under these conditions, these NPs are expected to be less stable compared to those synthesized with a thermodynamic control (heating at high temperature) that reach a global minimum in the Gibbs free energy. Okamoto et al. observed an Au-Pd alloy formation in face-centred cubic (*fcc*) structure at high temperatures up to the melting point for Au (1064 °C) for the whole Pd composition range.[39] In this sense, Wu et al. also observed the formation of an alloyed NP upon in-situ heating of a core@shell Au@Pd nanostructure up to 600°C. They showed that the evolution from a core@shell to an alloyed structure occurs through different steps, *i.e.* surface re-faceting, particle re-sphering and complete alloying. However, no phase segregation or phase separation was observed.[40] Prevot et al. studied the annealing effect of Au-Pd NPs deposited on amorphous carbon at 600°C and observed the bi-modal distribution of nanoparticles upon annealing, where small and big NPs became Au-rich and Pd-rich, respectively.[41] Nevertheless, their experiment resulted in the formation of completely separated Au and Pd NPs, which was explained by an Ostwald ripening mechanism.[41]

Here, we show the generation of phase-separated Au-Pd bi-metallic Janus NPs from an initial Au nanotriangle (AuNT)@Pd core@shell system with three different Pd loadings at high temperature (>900°C) via an in-situ annealing transmission electron microscopy (TEM) study under high-vacuum conditions. We have chosen AuNT as the core for our system as it contains very distinct plasmonic modes at its sharp corners and edges, which can be tuned very efficiently by a simple wet chemical method.[14,42,43] Also, large-scale production of AuNT can be easily yielded through a wet chemical process.[14,42] We have observed through extensive aberration-corrected high-resolution (scanning) TEM (HR(S)TEM) imaging and energy-dispersive X-ray spectroscopy (EDS)-STEM measurements that the core@shell Au@Pd nanostructure forms an Au-Pd alloy during in-situ heating at 400°C. This alloy nanostructure is highly stable up to 800°C, which is an expected behaviour as Au-Pd is predicted to be miscible in all composition and temperature ranges. Upon further heating to the 900°C-1000°C temperature range, the Au and Pd phases surprisingly got separated from each other and the separation is observed to be prominent for initial (AuNT)@Pd core@shell NPs with higher Pd loading. The Janus nanostructures, being produced by a heat treatment, are stable at ambient conditions for over one year.

## 2. Results and Discussions:

### 2.1. Initial TEM and UV-Vis study

Four specimens have been analysed in this work: bare Au nanotriangles (AuNT) and three core@shell Au NT@Pd nanostructures with different Pd loadings. A schematic of the formation process of AuNT and other three systems with different Pd loadings, which are subjected to in-situ annealing process is shown in Figure 1.

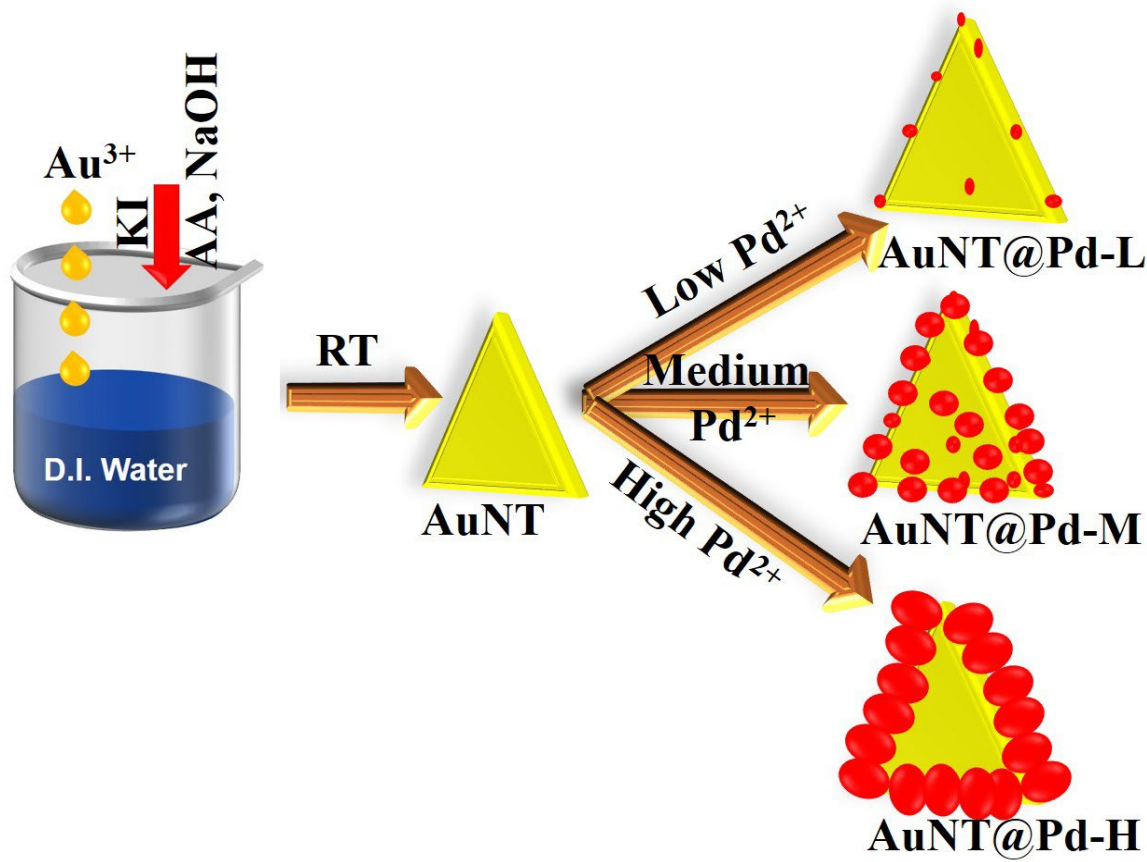

**Figure 1**. Schematic visualization of the four systems used for the in-situ study along with formation of initial AuNT core. Low, medium and high Pd-doped samples are indicated as AuNT@Pd-L, AuNT@Pd-M and AuNT@Pd-H, respectively. Potassium iodide, ascorbic acid and sodium hydroxide are indicated as KI, AA and NaOH, respectively.

Extensive TEM studies have been performed using aberration-corrected TEM to understand the initial structure of these specimens and, specifically, the Au–Pd interface in dependence on the Pd shell thickness. Figure 2a along with Figure S1, Supporting Information shows low magnification TEM images of the AuNT produced by the seedless method (Experimental section) along with the particle size distribution (Figure S1d, Supporting Information). The average side length of the equilateral AuNT is 73.11 ± 7.21 nm. The selected area electron diffraction (SAED) pattern (Figure 2b) taken along [-1 1 1] zone axis (ZA) shows that the AuNT are single-crystalline in nature. It also shows the presence of forbidden 1/3 (422) diffraction spots (red circles in Figure 2b) that indicate the formation of twin planes parallel to the (111) surface and perpendicular to the electron beam. These spots are forbidden in single crystal *fcc* structure but they can appear when there is a formation of twin planes parallel to each other, which has been observed previously for Au and Ag nanostructures bounded by atomically flat top and bottom surfaces.[44,45] Figure 2c shows a high-resolution TEM (HRTEM) image from the middle region of the AuNT marked by a red square in the inset of Figure 2b. The inset shows the fast Fourier transform (FFT) pattern taken from this region along [-111] ZA, which shows the presence of (202) and 1/3 (422) planes of Au. The inverse fast Fourier transforms (IFFT) calculated by selecting the (022) and 1/3 (422) frequency points are shown in Figure 2d and Figure 2e, respectively. They reveal the formation of stacking faults and line dislocations in the nanostructure. The presence of stacking faults is expected to enhance the alloying process during annealing, as it lowers the barrier for solute atom migration.[46–48]

According to an X-ray photoelectron spectroscopy (XPS) analysis (Figure S2, Supporting Information), the Pd atomic percentage in the different samples was 20 at%, 38 at% and 76 at%, respectively. However, as XPS is sensitive to the sample surface and Pd is located at the shell of the nanostructure, this quantification can overestimate the actual Pd percentage in the system. A better average composition of the different samples is obtained by inductively coupled plasma - optical emission spectrometry (ICP-OES) and EDS-STEM analyses of large clusters of the NPs. The ICP-OES analysis reveals a Pd concentration of 13.9 at%, 37.9 at% and 48.3 at% for the three Pd-doped samples, respectively. The EDS-STEM analyses are shown in Figures S3, Supporting Information and yield an average Pd doping of 13.6 ± 3.3 at%, 36.3 ± 3.7 at% and 47.2 ± 4.2 at%, which corroborates the ICP-OES measurements. For simplicity, from here on, we will name the three Pd-doped AuNT NPs according to their amount of initial Pd loading (low, medium and high) as AuNT@Pd-L, AuNT@Pd-M and AuNT@PD-H, respectively.

The Pd deposition process was initially studied using UV-Vis spectroscopy. The UV-Vis spectrum of bare AuNT shows a significant reduction of the 530 nm peak, which is characteristic of spherical Au NPs, indicating predominantly AuNT formation (Figure 2f). Bare AuNT show a plasmon peak at 670 nm wavelength. A blue shift and broadening of the plasmonic responses can be observed with increasing Pd deposition. The blue shift is attributed to the negative dielectric constant of Pd with respect to Au in the UV-Vis region. The broadening is due to the higher value of the imaginary part of the dielectric constant of Pd with respect to Au in the UV-Vis region.[14,43] The broadening and blue shift of the extinction peak indicates the formation of Au core- Pd shell nanostructures,[14,49–52] which is further confirmed by HRTEM and EDS-STEM measurements as shown in Figure 3 and in Figure S4.

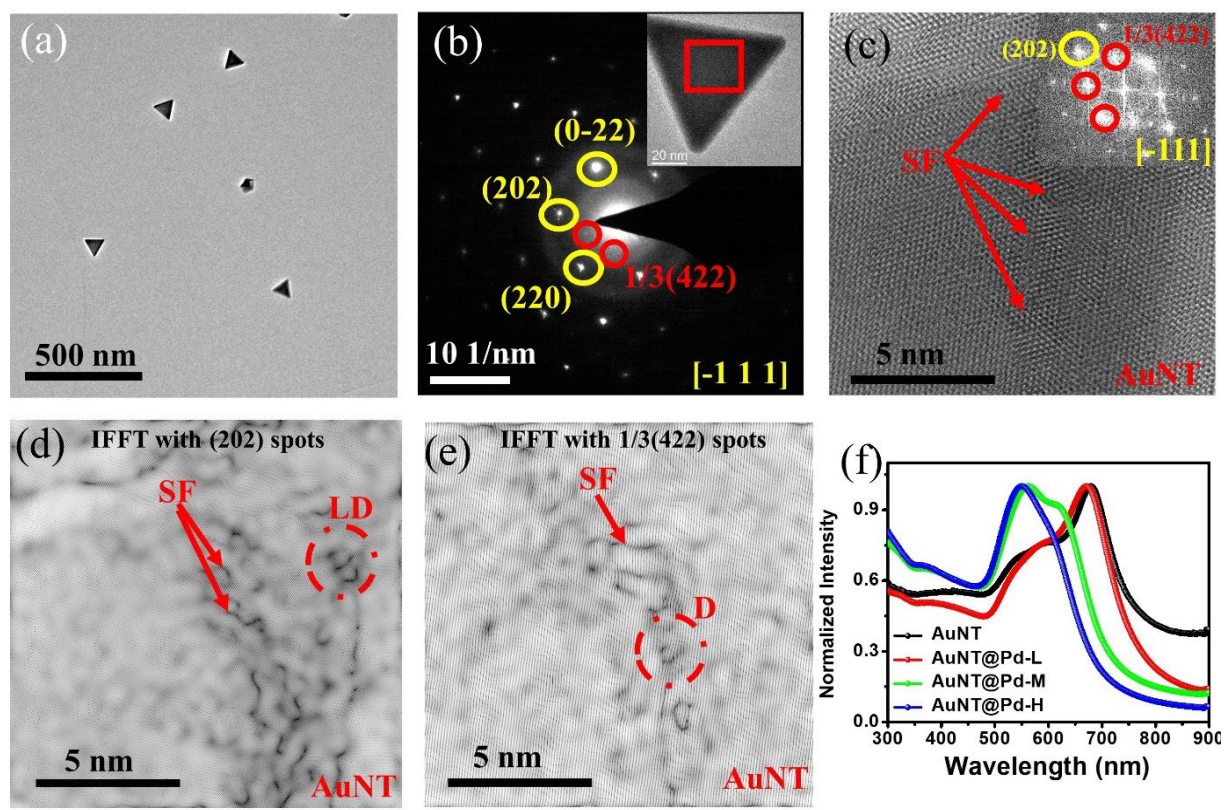

**Figure 2.** (a) Low magnification TEM image of AuNT. (b) Selected area electron diffraction (SAED) pattern taken along [-111] zone axis from the selected AuNT (shown in the inset). (c) HRTEM image taken from the flat surface of the AuNT marked by red square in the inset of Figure 2b. Inset shows the FFT pattern taken from this region showing presence of (022), and forbidden 1/3 (422) planes of Au. (d) and (e) show IFFT image of the region shown in c, by selecting (022) and 1/3 (422) frequency spots, respectively. Both the images show presence of defects i.e. stacking faults (SF; marked by red arrow) and dislocations (LD; marked by red circle) on the AuNT. (f) UV-Visible extinction spectra (normalized) of bare AuNT and Pd doped AuNT showing a clear blue shift and increased linewidth with increasing the Pd content.

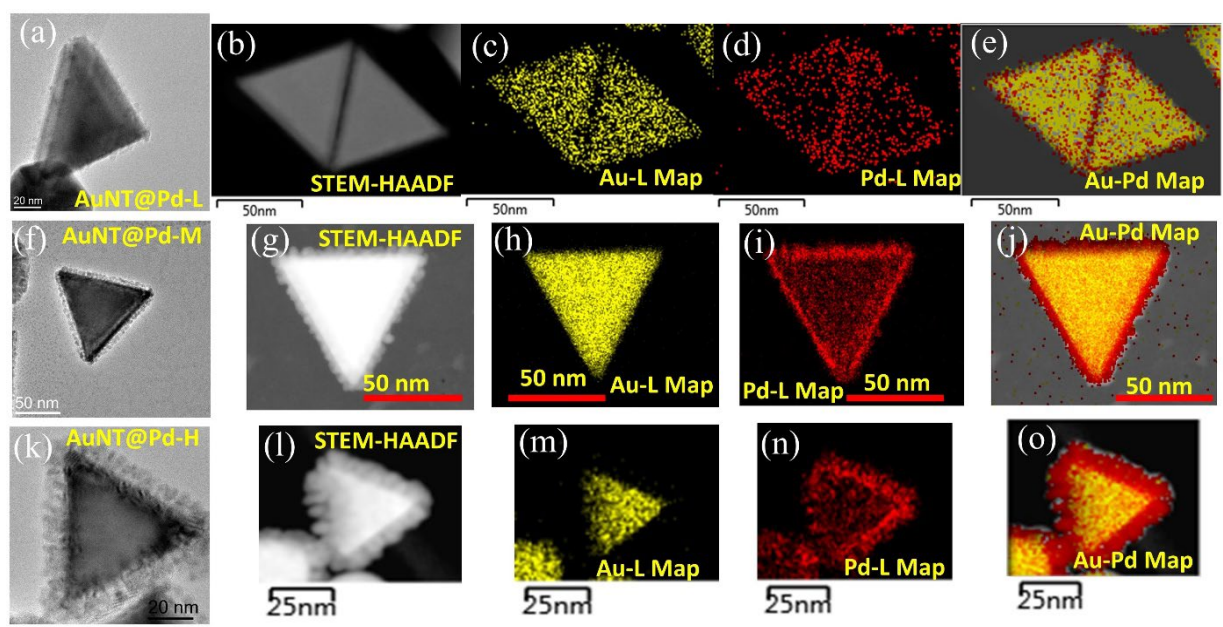

**Figure 3.** EDS analyses of the initial AuNT for different Pd loadings of (a-e) AuNT@Pd-L, (f-j) AuNT@Pd-M and (k-o) AuNT@Pd-H. (a), (f) and (k) show BF-TEM images. (b), (g), (l) show STEM-HAADF images of three Pd deposited AuNT analysed by EDS, which yielded (c), (h), (m) Au-L maps (d), (i), (n) Pd-L maps and (e), (j) and (o) shows the Au-Pd combined maps, respectively.

Figure 3 shows example nanostructures for the different Pd-deposited AuNT samples and their EDS elemental mapping to understand the spatial distribution of Au and Pd within the NP. The Pd thickness is varied such that the Pd-doped AuNT shows significant differences in their optical property compared to each other, which tailors their response when they are photothermally activated and provides an idea of the approximate thickness of the Pd layer.

Figure 3a shows a TEM image of the AuNT@Pd-L sample. An HRTEM image of this sample is displayed in Figure S4a, Supporting Information. It can be seen that the Pd took the form of islands instead of growing as a film. This indicates that Pd atoms prefer to grow or stick to other Pd atoms compared to the Au surface, resulting in this 3D island kind of growth of Pd on AuNT. Figures 3 (b-e) show a STEM high-angle annular dark-field (HAADF) image, the Au-L EDS map, the Pd-L EDS map and the Au-Pd combined maps, respectively. The predominant presence of Pd at the shell region of the NPs is clearly evidenced.

Figure 3f shows a TEM image of the AuNT@Pd-M sample. A complementary HRTEM image of this nanostructure is displayed in Figure S4b, Supporting Information. The average thickness of the grown Pd nanostructure is 4.9±0.8 nm, as observed from different HRTEM images. The SAED pattern displayed in the inset of Figure S4b shows the presence of Au (220) and Pd (220) planes, respectively. The EDS measurements from this nanostructure (Figure 3g-j) showed that the thickness of the Pd layer increases compared to the AuNT@Pd-L sample and that, again, the majority of the Pd atoms reside at the shell region. Figure 3k and Figure S4c (Supporting Information) correspond to the TEM image of the AuNT@Pd-H sample and its HRTEM image, respectively. The Pd shell is observed to grow in a columnar-like structure. The SAED pattern taken from the NP showed diffraction spots corresponding to Au (111), (200) and (220) planes and Pd (111) and (220) planes, respectively. The average height of the Pd columnar structure is 7.12 ± 0.97 nm. Figures 3 (l)-(o) show STEM-HAADF image, Au-L map, Pd-L map and Au-Pd combined maps, respectively and it shows an average overall content of 57.3 at% Pd. This is considerably smaller than the value obtained from XPS (76 at% Pd) but slightly higher compared to the EDS and ICP-OES analyses. The Pd loading is significantly higher in this sample compared to the AuNT@Pd-L and AuNT@Pd-M samples.

In summary, four samples are synthesized consisting of bare AuNT, 13.9 at% Pd doped AuNT (denoted as AuNT@Pd-L), 37.9 at% Pd doped AuNT (denoted as AuNT@Pd-M) and 48.3 at% Pd doped AuNT (denoted as AuNT@Pd-H), which are subjected to in-situ thermal annealing

TEM up to 1000°C to study how the structure evolves with temperature and what is the role played by the Pd layer in the stability of the nanostructure.

## 2.2. In-situ heating TEM studies

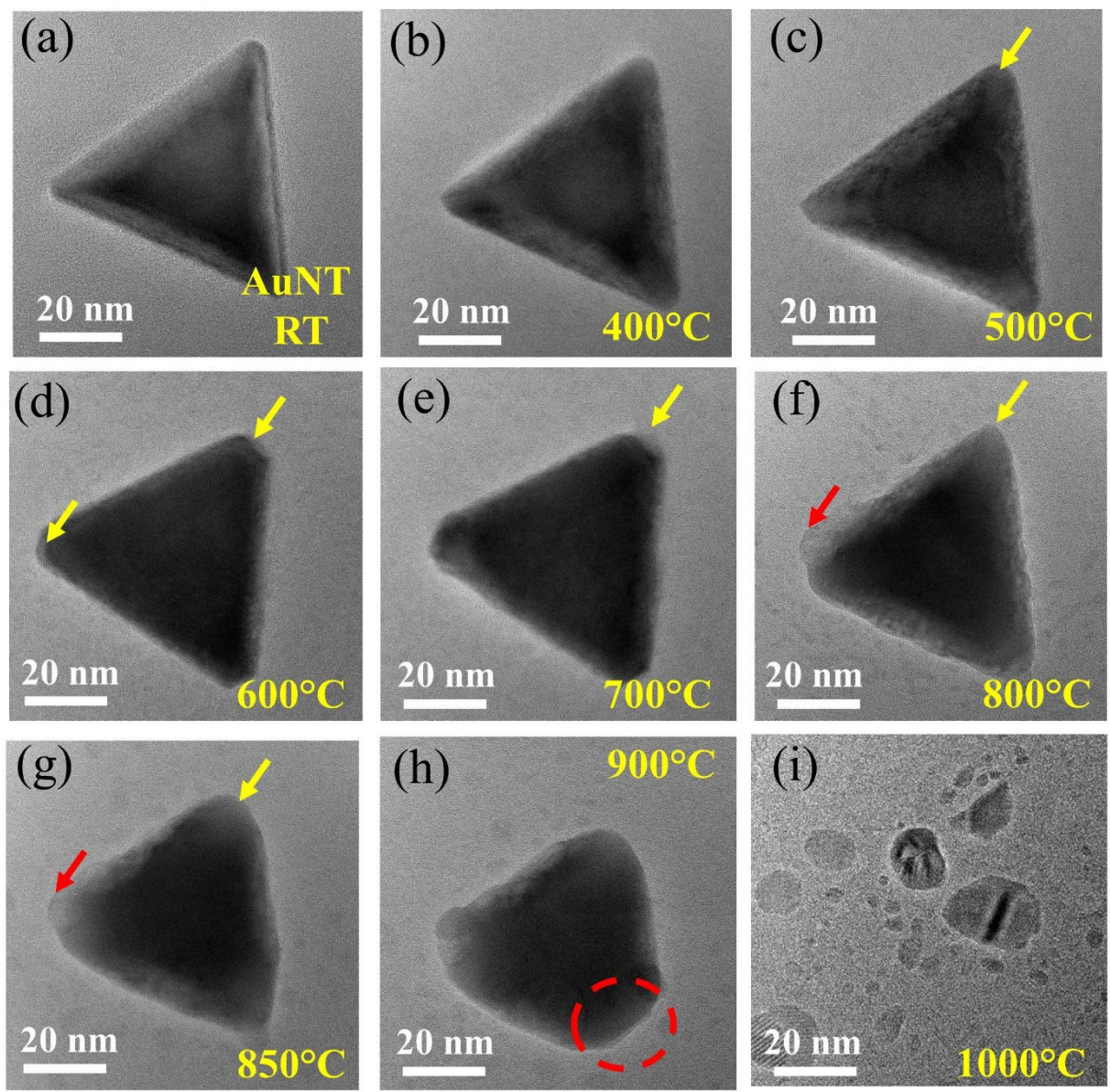

**Figure 4.** Structural variation of bare AuNT: (a) RT and during in-situ annealing at: (b) 400°C, (c) 500°C, (d) 600°C, (e) 700°C, (f) 800°C, (g) 850°C, (h) 900°C, and at (i) 1000°C, respectively. The yellow arrows indicate points from where the particle decomposes and the red arrow shows the points where small Au NPs get re-deposited on the AuNT due to an Ostwald ripening process.

Figure 4 shows the structural variation of a single AuNT nanostructure with in-situ annealing at different temperatures. The images were taken after keeping the NP at each temperature for 20 minutes.

For pure AuNT, the initial deformation of the particle is observed to form at the tips, where the surface energy is higher (marked by yellow arrows). However, at higher temperatures (around 600 °C), some small particles, which are formed from the decomposed AuNT, again attach to the AuNT (marked by the red arrow in Figure 4f) due to an Ostwald ripening process.[53–55] The AuNT deforms mostly from one tip, which continues going blunt and the particle finally melts at 1000°C (Video S1, Supporting Information). Figures S5 and S6 (Supporting Information) show the HRTEM and FFT analyses conducted to determine the interplanar spacing from the outer and inner portions of the AuNT at different temperatures. Therefore, FFT patterns (Figure S6) are obtained from different regions marked by red and yellow squares in the AuNT (Figure S5) and the distance between the (220) spots is measured for all the NPs (AuNT and Pd doped AuNT) and at all temperatures, but for convenience, only exemplary images of AuNT at RT and 900°C are shown. The detailed variation of the (220) lattice parameter with temperature for all the NPs will be discussed later. The variation of the SAED pattern with temperature is shown in Figure S7 (Supporting Information). The molten phase at 1000ºC leads to strong diffraction rings (Figure S7, Supporting Information), which however already start appearing from 600ºC and gain in intensity with increasing temperature (Figure S7d-g, Supporting Information). Though bulk Au has a melting point of 1064°C, the nanostructures use to have a smaller melting point compared to their bulk counterparts, which is attributed to the higher surface-to-volume ratio of the nanostructure.[56]

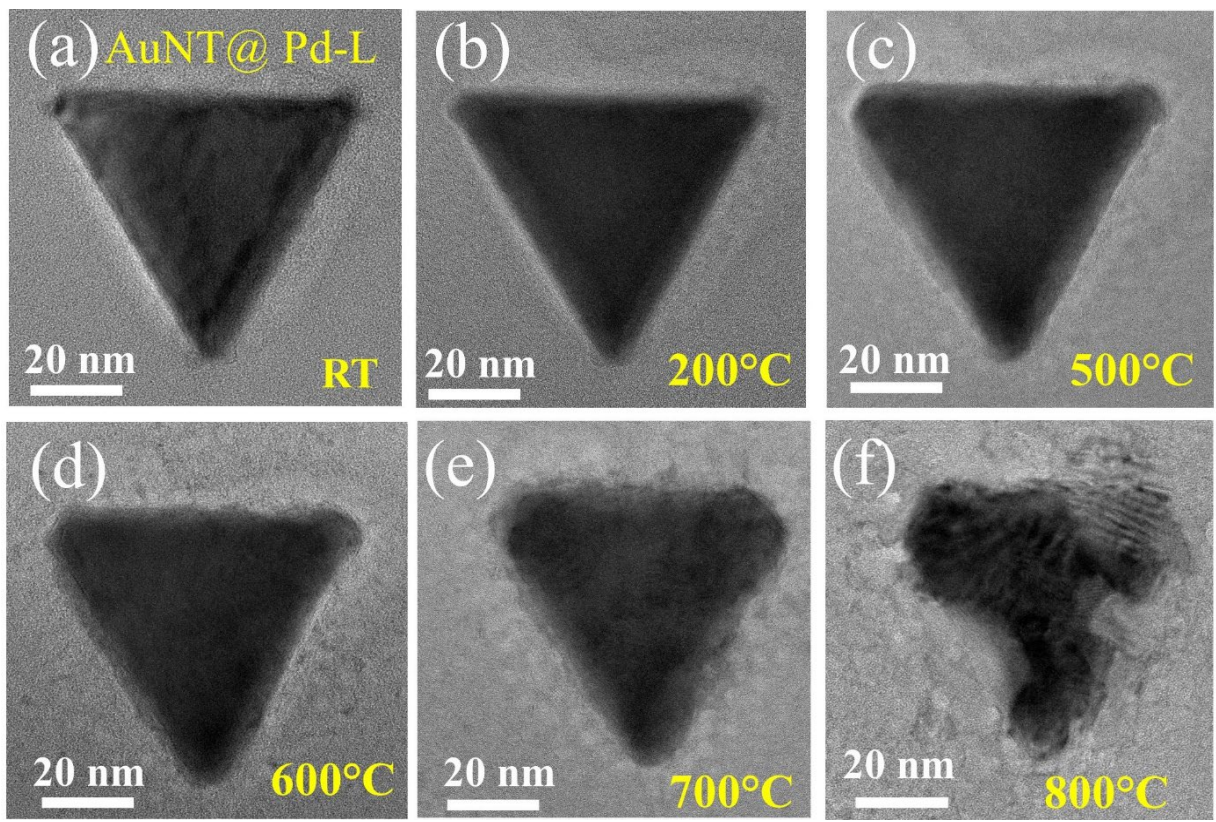

**Figure 5.** Evolution of AuNT@Pd-L sample at: (a) RT (b) 200°C, (c) 500°C, (d) 600°C, (e) 700°C, (f) 800°C, respectively.

Figure 5 shows the evolution of an AuNT@Pd-L NP with increasing temperature. Similar to the case of pure AuNT, the deformation starts again from the corner of the NP. However, unlike the pure AuNT, which melts at 1000°C, this sample shows melting behaviour at only 800°C. This clearly indicates that a small Pd coating indeed adversely affects the stability of the structure. Figure S8 (Supporting Information) shows the different regions of the inner and outer portions of the NP, marked by yellow and red square regions, respectively, which are used to obtain the FFT pattern for measurement of the variation of lattice parameter upon in-situ TEM heating.

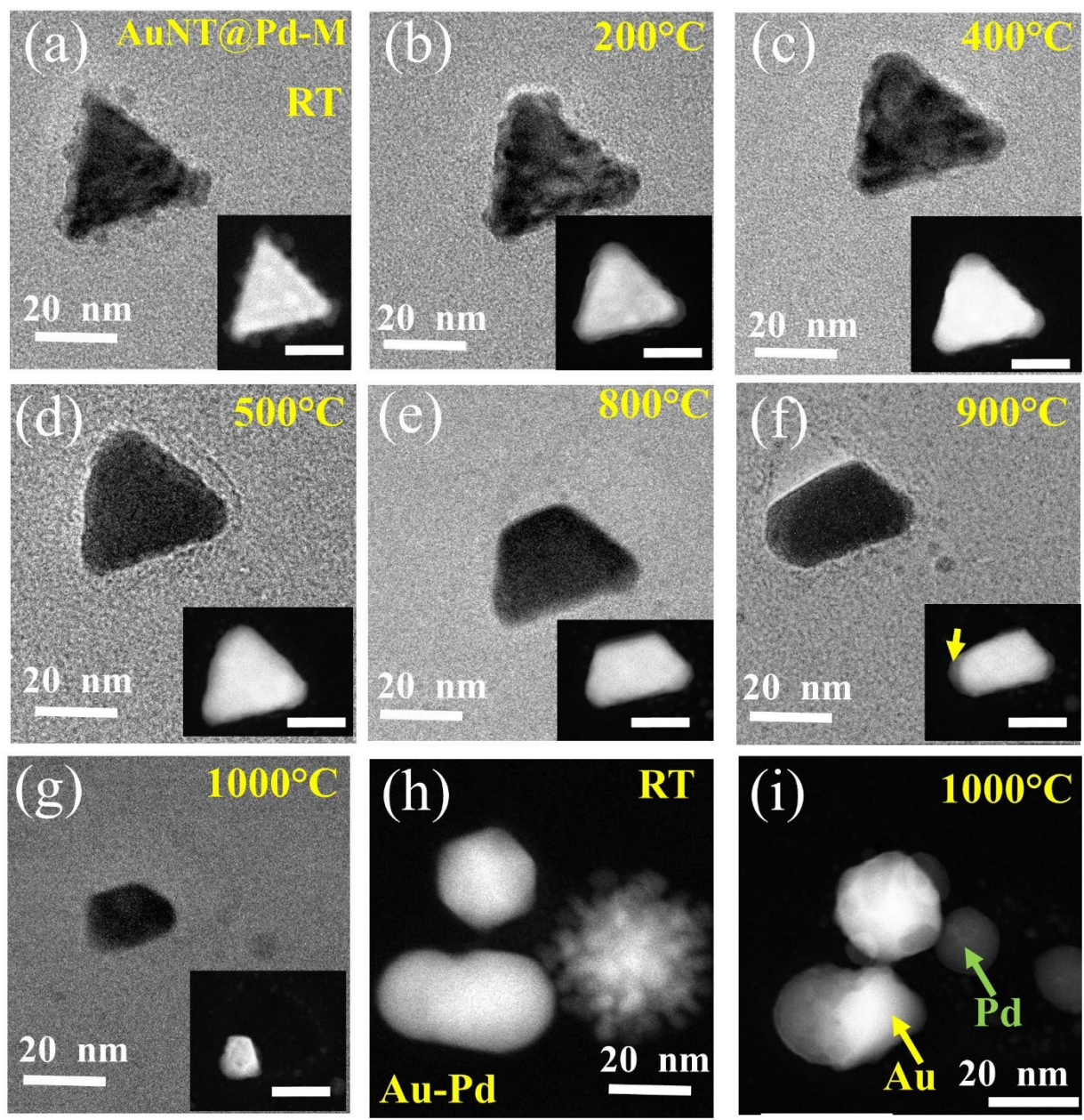

**Figure 6.** BF-TEM and STEM-HAADF (inset, scale bar 20 nm) images of the evolution of AuNT@Pd-M NP at: (a) RT, (b) 200°C, (c) 400°C, (d) 500°C, (e) 800°C, (f) 900°C, (g) 1000°C, respectively. Figures (h) and (i) show the STEM-HAADF images of irregular shaped Au-Pd NP at RT and at 1000°C, respectively.

Figure 6 shows BF-TEM and STEM-HAADF images (inset) of an AuNT@Pd-M NP and the variation of the nanostructure with temperature. The initial structure at RT comprises Pd islands on the AuNT as shown in Figure 6a. However, during in-situ heating at 200°C, the island-like structure immediately disappears, indicating a Pd-Au alloy formation. The alloyed structure is confirmed via EDS spectroscopy after heating the structure to 500°C, as shown in Figures S9a-S9d (Supporting Information). The EDS maps show that Pd and Au are present uniformly over the entire NP and not in a core@shell configuration as observed initially at room temperature (RT). The particle shows Au and Pd atomic percentages of 58.76% and 41.24%, respectively, and the Pd content is larger than the average values obtained by the ICP and EDS measurements. Upon further heating, one tip of the nanostructure is decomposed at 800°C. However, a Pd-dominated region is obtained after heating the NP to 900°C for 20 minutes, which is shown by the yellow arrow in the inset of Figure 6f. The EDS spectra taken from this Pd-dominated region reveal a Pd concentration of 52 at%, compared to only 15 at% Pd in the overall NP, as shown in Figure S9e-S9h (Supporting Information). At 1000°C, the particle was totally decomposed, leaving behind small NPs as shown in Figure 6g. Figure 6h and Figure 6i show the comparison of irregularly shaped Au-Pd NP before and after annealing to 1000ºC. The Au-Pd separation also occurs in this case, indicating that the phase separation phenomenon is not specific for the triangular structure, but a general feature of the Au-Pd alloy at high temperatures. It also proves that the separation is not induced by the electron beam, as those NPs were not exposed to the electron beam during the in-situ heating process.

Figure S10 (Supporting Information) shows the formation pathway towards the regions with different Pd and Au compositions. It can be seen that during heating at 900°C, the Pd and Au regions decompose separately (indicated by the separate movement of the Au and Pd region in Video S2 (Supporting Information)) and remain stable at 900°C and also when the temperature is reduced to 400°C for performing an EDS measurement. This is unique, considering the fact that Au and Pd are completely miscible on the bulk scale at all temperatures and for the whole composition range. Further heating of the NP to 1000°C leads to a subsequent melting of the NP, which is shown in Figure S11 (Supporting Information) and Video S3 (Supporting Information). This shows that, similar to the bare AuNT, the melting process starts from the surface or the tip region of the NP and proceeds towards the centre. Different theoretical studies on bi-metallic Au-Pd/Au-Pt NPs based on molecular dynamics (MD) simulations showed that the melting of the alloy nanostructure occurs through two phases: one pre-melting phase, during which the surface begins to melt initially, and as the heating progresses, the entire nanoparticle transitions into a molten state, a behaviour consistent with our observations.[57–60]

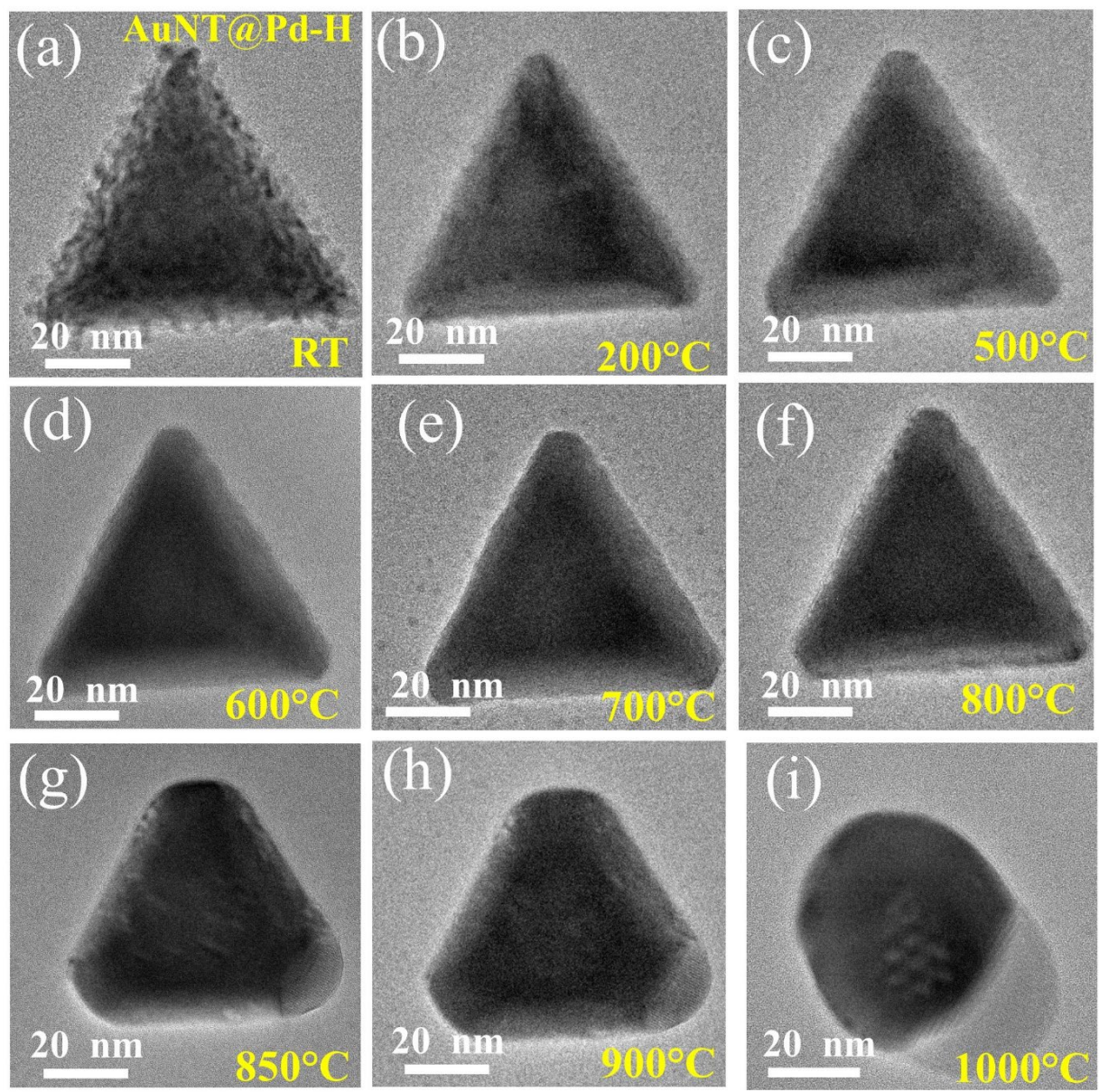

**Figure 7.** Evolution of AuNT@Pd-H NP at: (a) RT, (b) 200°C, (c) 500°C, (d) 600°C, (e) 700°C, (f) 800°C, (g) 850°C, (h) 900°C, and (i) 1000°C, respectively.

To confirm the Au-Pd separation, and also to understand the role of Pd thickness on the thermal stability of the NP, a similar in-situ TEM heating experiment of the AuNT@Pd-H sample is done. Figure 7 and S12 show TEM and HRTEM images of the structural variation of the nanostructure with in-situ heating, respectively. The Pd columnar structure immediately gets lost with heating to 200°C as shown in Figure 7b and Figure S12b (Supporting Information). With further heating, the structure shows a similar behaviour to the previous NPs, i.e., it shows the formation of an Au-Pd alloy. However, at 1000°C, remarkably, Au and Pd show a clear separation (Video S4 (Supporting Information)). Figure S13 (Supporting Information) displays individual images/frames taken from Video S4 (Supporting Information) during different time intervals. They reveal that the separation process starts from one corner and that the Au portion gradually moves inwards (towards the centre), leaving behind a thin Pd outer layer. We also observe an alternating contrast (or perturbation behaviour) around the central region of the NP (Figure 8a), which seems to be separated from the surface region by a tilted interface between the Au and Pd phases. This indicates a reduced involvement of the surface region in this process. The FFT pattern (Figure 8b), taken from the Pd region in Figure 8a, shows Pd (111) FFT spots. Also, Figure 8a displays that the interface between the Au and Pd region is not sharp but contains a Moiré pattern due to the overlap of Au and Pd interplanar spacing indicating a separation of Pd and Au phases in vertical direction due to the tilted phase boundary. The HRTEM images taken from the Au region show that the crystallinity at the central region of the NP close to the interface decreases, whereas the surface region shows good crystallinity compared to the core region and still consists of Au-Pd alloy as seen from the Wien-filtered HRTEM image of the surface shown in Figure 8d.[52] This again proves the reduced involvement of the surface region in the separation process. The SAED pattern obtained from the NP after heating also shows reduced crystallinity (molten phase) indicated by a ring pattern (Figure S14b, Supporting Information).

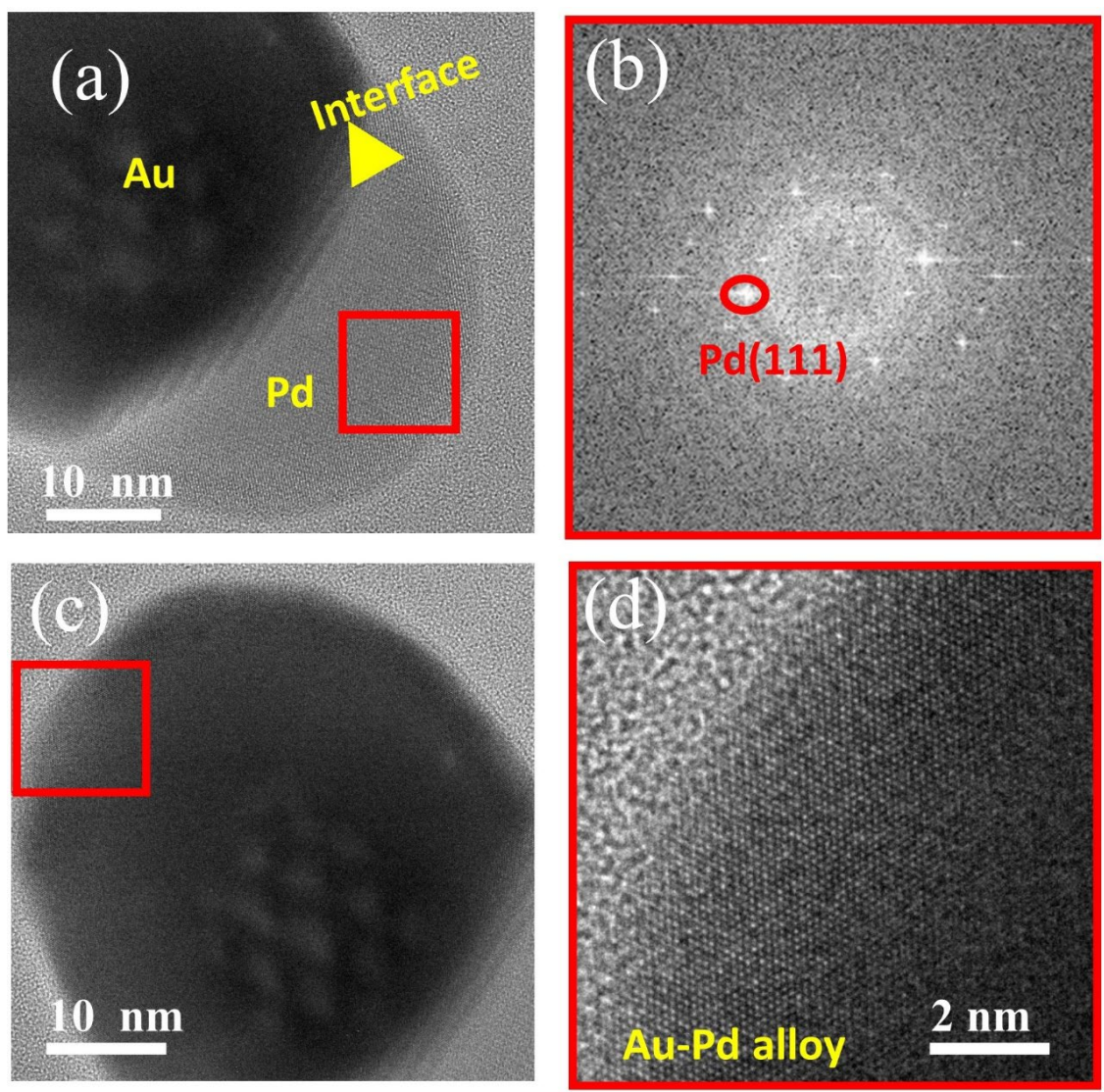

**Figure 8.** (a) HRTEM image of the nanostructure after annealing at 1000°C. (b) FFT image from the region marked by red square in Figure a, showing Pd (111) FFT spots. (c) HRTEM image of the Au portion and (d) Wien filtered TEM image of the surface marked by red square in Figure 8c showing clear atomic contrast, which indicates the stability of the Au-Pd alloy at the surface.

The same Au-Pd separation phenomenon was observed for several other particles shown in Figure S15 (Supporting Information), where six different particles (not exposed to the electron beam during in-situ heating) show the same evolution. To further confirm that the separated portion is Pd, EDS-STEM analyses were performed on these NPs at RT.

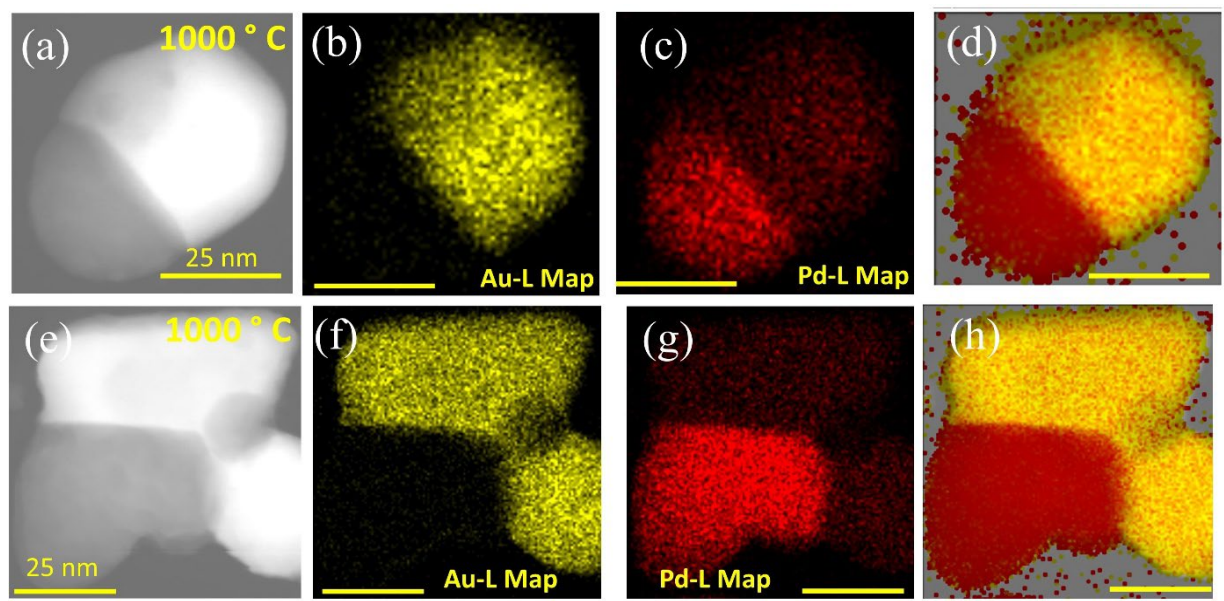

**Figure 9.** EDS elemental mapping of AuNT@Pd-H for two different NPs after in-situ heating and keeping the sample in ambient atmosphere for one week at RT. Figures 9 (a) and (e) show STEM-HAADF images on which EDS analysis has been conducted. (b) and (f) correspond to Au-L maps, (c) and (g) Pd-L maps, (d) and (h) Au-Pd combined maps, respectively, which clearly show the Au-Pd phase separation for both NPs.

The phase-separated NPs exhibit remarkable stability at room temperature under ambient conditions, indicating a stable structure. The EDS elemental maps obtained from the NPs clearly show the separation of Au and Pd, as depicted in Figure 9. Further EDS-STEM analyses of additional phase-separated nanostructures (Figure S17, Supplementary Information) confirm this observation. The data also indicates the presence of Pd in some amount on the Au side, but no detectable Au on the Pd side. Although the initial nanostructure contains around 48 at% Pd, the Janus NP contains only 32 at% Pd. The line profile obtained from the Janus NP shows a distinct dip in the Pd concentration around the mid-region, as shown in the line profile in Figure S17c (Supporting Information), while some Pd concentration is observable on the surface of the Au side. This aligns with the observation seen in Figure 8d. The same phenomenon is also observed from the EDS line spectrum obtained from another particle, which exhibits a dip in the Pd concentration around the mid-region, while the Pd concentration increases at the surface (Figure S17f, Supporting Information). Additionally, the in-situ TEM study clearly indicates that the structural stability of the alloyed nanostructure increases with an increase in the amount of Pd. For instance, AuNT@Pd-L shows a melting behaviour at 800°C, AuNT@Pd-M exhibits separation behaviour at 900°C and melting behaviour at 1000°C, and AuNT@Pd-H displays a more prominent separation behaviour at 1000°C. This is consistent with previous theoretical studies based on MD simulations.[57–60]

It is important to note that the NPs selected for these in-situ measurements were similar in size except for the AuNT@Pd-M one, which shows a side length of only 39.97 ± 1.28 nm. The other NPs tracked during these in-situ measurements (bare AuNT, AuNT-Pd@L and AuNR@Pd-M) have an almost identical side length of 63.60 ± 0.94 nm, 65.05 ± 0.69 nm and 65.06 ± 0.58 nm, respectively (Figure S18, Supporting Information). In general, the nanoparticle melting point decreases with decreasing the particle size due to the relative increment in the number of surface atoms compared to the atoms in the bulk.[56–58] Here also, we observed that the smaller AuNT@Pd-M showed an early onset of phase segregation behaviour starting at 900°C (Figure 6f), whereas, the larger AuNT shown in Figure 7f displays the complete phase segregation behaviour at 1000°C. The smaller NP also melted at 1000°C, whereas the highest Pd-loaded AuNT shows the surface pre-melting process followed by phase separation at this temperature.

To verify the reproducibility of the Janus NPs' formation and their stability, we have studied the in-situ chips containing the annealed NPs after keeping them at ambient conditions for over a year. The images along with EDS mappings are shown in Figure S19, Figure S20, and Figure S21, Supporting Information and reveal a large amount of Janus NP formation. A part of the NPs are found in clusters, showing a clear separation between separate Au and Pd NPs as shown in Figure S16, Supporting Information. The size of the Janus NPs is found to vary and is attributed to the density of the NPs on the support film. NPs surrounded by a larger number of additional NPs are more probable to grow due to an Ostwald ripening process during the phase separation process. A detailed explanation of the phase separation mechanism is elaborated below, in the discussion section of the manuscript.

To summarize the experimental results of the in-situ heating study, Figure 10 shows a schematic description of the thermal evolution of bare AuNT and Pd-doped AuNT and how the final structure varies with the initial Pd concentration. The bare AuNT NPs melt at 1000°C into a small number of NPs. Video S1 depicts the detailed evolution of the NP just before melting. It clearly shows an initial decrease of the sharpness at the edges. With further heating, at the start of the melting process, the particle gets thinner initially due to loss of material and after that, the decomposition starts from one side and finally produces smaller NPs. For all the Pd deposition cases, an Au-Pd alloy is produced, and this Au-Pd alloyed nanostructure is stable in the temperature range of 400 -700°C. The addition of a low amount of Pd coating results in a decrease of the decomposition temperature to 800ºC. However, with increasing Pd concentration, the melting temperature increases again, and regions dominated by Au and Pd, respectively, are obtained. As it can be expected, the size of the Pd region depends on the initial concentration of the Pd doping. Higher Pd amount leads to an increase in the area dominated by Pd.

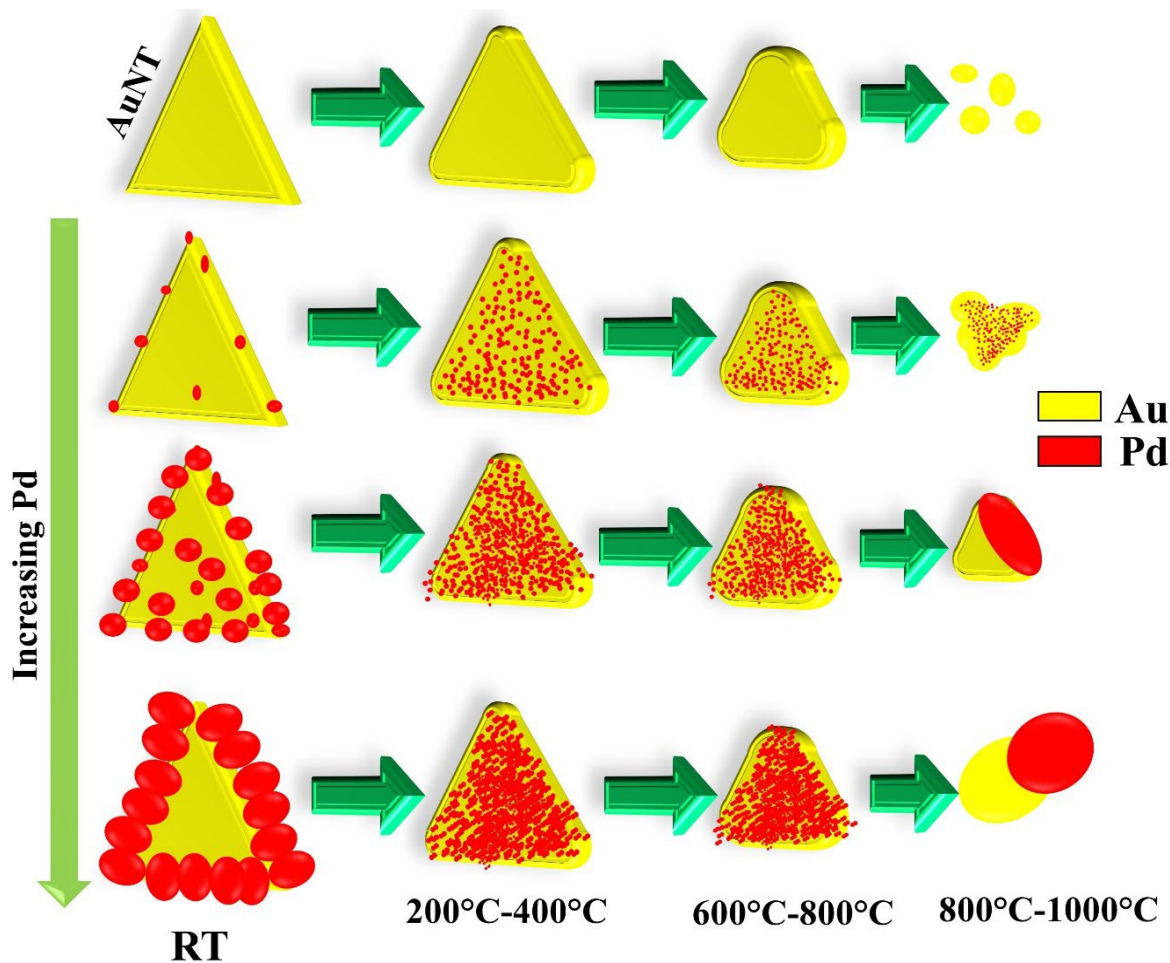

**Figure 10.** Schematic of thermal evolution of bare AuNT and Pd deposited AuNT.

The changes in the lattice parameter from the inner and outer portions of pure AuNT and Pd-doped AuNT are tracked using FFT patterns obtained from these portions and the evolution is shown in Figure 11. This indicates that the interplanar spacing within the core is 1.44 Å, which corresponds to the Au (220) plane. The core area shows very good stability in terms of lattice parameter for bare AuNT and Pd-doped AuNT and is plotted in Figure 11a. Figure 11b shows the variation of the lattice parameter of the shell region of the nanostructure. At lower temperatures, the lattice parameter of Pd deposited AuNT shows a lower interplanar spacing compared to bare AuNT, which is due to the smaller distance of the Pd (220) plane. However, as the temperature increases, the interplanar spacing in the shell increases and indicates the formation of an AuPd (220) alloy at the surface. The interplanar distance between (220) planes of bare AuNT of the shell region shows a large increment during annealing at high temperature (Figure 11b), possibly due to high thermal vibration during melting[61,62] contributed by the higher surface energy in these positions due to rich presence of lower co-ordinated surface atoms. Whereas, Pd deposited AuNT shows less increment of the (220) interplanar distance at high temperature, indicating that Pd incorporation provides increased structural stability upon annealing.

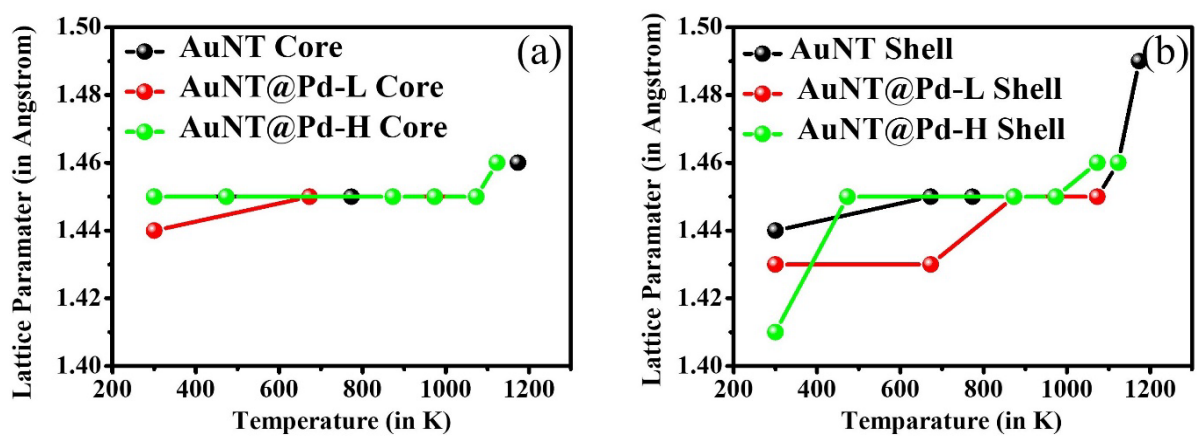

**Figure 11.** Variation of (220) interplanar spacing for bare AuNT and Pd-doped AuNT for the (a) core and (b) shell region, respectively.

## 3. Discussions

To interpret our observed experimental results and, specifically, give an explanation for the Au-Pd phase separation, we review previous works on related bi-metallic nanostructures. Braidy et al. [63] studied the thermal annealing effect on Au@**Pt** core@shell NPs and observed the same phenomenon of an Au-**Pt** phase separation by a clear interface at high temperature (800°C) and inter-diffusion forming an alloy structure at intermediate temperatures. As Au and Pt have a broad miscibility gap in their phase diagram, the phase separation was expected. Using Buttler's equation to derive the surface energy for components with strongly differing surface energy values, they obtained that the mixing enthalpy for Au-Pt shows a large positive value at high temperature and high Pt loading. Such phase-separation equilibrium is also seen for related bi-metallic and ternary alloy systems, i.e. Cu-Ni,[64] Au-Ni,[65] Au-Pt,[66] Au-Ag-Cu,[67] Ag-Cu[68] and Ag-Au, Ni-Pt.[69] In all those reports, the researchers concluded that the phase diagram of the nanoscale alloys is much more complex than their bulk counterparts because of different factors, i.e., high surface-to-volume ratio, segregation energy as a function of the distance from the surface layer, interactions within the local atomic environment, and configurational aspects, namely, the relative number of sites in each region. Knowing the overall contributions from these factors, one can further apply Monte-Carlo (MC) or MD simulations, or both, to predict the nano-alloy morphology in different conditions.

For Au-Pd nanoscale systems, such a phase-separation behaviour resulting in the formation of two different hemispheres of Au and Pd and thus, generating a hybrid Janus NP was not previously reported with annealing. Indeed, until now, only the formation of Au-Pd alloy[70,71] and/or complete separation of Au and Pd in separate NPs providing a bi-modal distribution have been found.[41,72,73] Fang *et al.* observed that upon annealing an Au@Pd core@shell system, the Pd atoms tend to move towards the core with increasing temperature for all coverages, indicating an alloy formation and the transition from a core@shell to a homogeneous AuPd alloy.[71] However, the tendency of Pd atoms to migrate towards the Au core is higher at low Pd loading. For higher Pd loadings, the migration tendency of Pd atoms is reduced, resulting in a higher concentration of Pd atoms at the surface. They also did not observe oxidation or formation of PdO on the surface, which makes the Au-Pd NP very resistant to oxidation at RT or above, here 300°C. As bare Pd NPs tend to become oxidized at high-temperature annealing, this indicates that Au also stabilizes metallic Pd atoms against oxidation. Confirming this observation, our study found no evidence of oxidation in the structures formed during the in-situ annealing. Indeed, HRTEM analysis suggests Pd (111) and not PdO (111) (Figure 8). Furthermore, detailed EDS analyses (Figures S22-S25, Supporting Information) do not reveal any evidence of oxygen incorporation into the nanoparticle even after keeping the sample in the ambient atmosphere for one week. We consider that the excellent chemical stability can be mainly attributed to the chemical inertness of Au, which was found to lead to superior chemical stability and oxidation resistance upon incorporation to Ag, Pd and $TiO_2$.[13,74–76] While Au provides the chemical stability, Pd is considered to provide thermal stability to the bi-metallic/alloyed system, which is due to the higher melting temperature of Pd compared to Au.[75] This synergistic effect makes this system a really stable one.

Simson et al. also observed the same movement of Pd NPs seen by Fang et al. while annealing at 160°C.[77] Also, they observed leaching of Pd NPs at low temperatures, which we have seen in our case from the in-situ annealing study, as proved by the presence of small Pd NPs in a larger area around the initial nanostructure after the study (Figure S10 (Supporting Information)). Vece et al. observed a compositional change in the behaviour of Au-Pd NPs upon hydrogenation, where the mobility of Pd and Au changed and Pd got detached from the Au-Pd system and produced large-size particles due to higher mobility.[72]

Berthier et al. used an effective site energy model to deduce the enthalpic and entropic contribution to the permutation-free energy and thereby theoretically determined the Au-Pd bulk phase diagram.[33] The authors obtained a miscibility gap for high Pd loading and beyond the order-disorder temperature.[33] Their consideration of coupling between chemical effect (mean field approximation using Ising model) and elastic effect (size effect) to determine the variation of chemical potential with Pd concentration leads to a size-dependant composition and finally to a miscibility gap on higher Pd concentration. Ter-Organessian et al. used the reactive force-field (ReaxFF) formalism to study the roles of composition, temperature, and NP size on phase separation, alloying, and surface segregation phenomena in NPs consisting of several thousand atoms.[34] They also observed that Au-Pd alloy shows a phase separation behaviour when smaller-sized NPs with high Pd content are annealed to high temperature, exactly similar to our observation. Prévot et al. experimentally observed a complete phase separation of amorphous C-supported Au-Pd NPs during annealing at 600°C giving rise to two distinct populations of NPs i.e. Au rich small NPs and Pd-rich large NPs.[41] Using kinetic

Monte-Carlo simulations, they observed that the generation of separate Au and Pd NPs occurs from the alloyed ones when the Pd atom's surface energy is 28 eV.nm$^{-2}$, much higher than Au (9.4 eV.nm$^{-2}$). However, a phase segregation is observed for a lower surface energy value of Pd, which is 12.7 eV.nm$^{-2}$. They also showed theoretically that, for low doping of Pd < 40%, the system gets unstable upon further attachment of Pd atoms (due to increment in chemical potential) and as a result, the Pd atoms dissociate from the particle. However, for higher Pd concentration, the chemical potential of the system decreases when Pd atoms get attached to it, indicating an Ostwald ripening process. We also observe the same phenomenon, where the initial AuNT@Pd-H NP contains 48.3 at% Pd at RT and the Janus nanostructure showed 32 at% Pd in the final configuration, indicating leaching of Pd atoms during heating. The leaching of Pd at lower annealing temperature indicates that the system energetically favours the detachment of Pd around 500-600°C. When the particle gets smaller in size, the reverse process occurs, the system reduces its chemical energy by attachment of Pd atoms. However, in contrast to the observation of Prévot et al., we did not obtain a complete separation of Au but obtained a phase separation between Au and Pd within a single nanostructure, indicating that the surface energy and mobility of Pd must be higher than for Au at high T,[41] but not high enough to totally separate it out from the NP. This finally produces a Janus nanostructure, which remains stable at RT.

All the above-mentioned observations provide us the necessary grounds to explain the phase-separation behaviour, which is discussed in detail for the AuNT@Pd-H NP in the following. At 850°C, the AuNT@Pd-H NP shows an initial surface pre-melting and forms an Au/Pd separation region as shown in Figure S26 and Video S5, Supporting Information. However, as Au has a lower surface energy than Pd at this configuration, the Au atoms come to the surface region of the NP (indicated by the darker contrast at the surface region in Figure S26b and Figure S26c, Supporting Information) and restart the surface pre-melting process and finally again produce an Au-Pd alloy structure in the corner. The interface is still there after annealing for almost 19 minutes (1120 seconds) indicated by the dark and white contrast in Figure S26c. The video and images at 900°C (Videos S6, Figure S26(d)-(f), Supporting Information) again show the surface pre-melting followed by the formation of an Au-Pd alloy at the corner. This process is occurring repeatedly, likely driven by the lower surface energy of Au compared to Pd for this size and Au/Pd ratio. Finally, at 1000°C, the phase-separation behaviour can be explained by the tendency of Pd atoms to segregate at the surface at high temperature (beyond the order-disorder transition temperature) and at higher Pd concentration. This is attributed to the high mobility and the high surface energy of Pd and possibly to the high value of mixing enthalpy in this configuration. The formation of phase-separated Au-Pd NPs is a process completely driven by thermodynamics, which can be divided into two parts. The first process is the pre-melting of the Au atoms at the sharper edge at high temperature (or at the temperature where phase separation occurs) due to the presence of atoms with a low coordination number atoms at the high curvatures at the edges. Subsequently the Au atoms move inwards building an interface as shown in Figure S13, Supporting Information. In the second step, when the Pd/Au ratio goes beyond a certain critical value (which also depends on the NP size), the Pd atom's mobility increases, and they move towards the surface to finally produce the Janus nanostructure and reduce the overall chemical potential of the system. This formation process agrees with the segregation rule stated by Gusbiers et. al., where, in the case of two metals with complete or partial miscibility, the element with higher melting temperature will segregate at the surface, here, Pd. This rule is seen to be applicable for a large number of bi-metallic systems.[78] As the system is driven by a thermodynamic process, the final nanostructure exhibits a global minimum in the Gibbs free energy, which agrees with the observed high stability of the phase-separated NP. We have observed the Au@Pd Janus nanostructures in large numbers even after keeping the samples in an ambient atmosphere for a year.

## 4. Conclusions

The structural variation of bare Au nanotriangles (AuNT) and Pd-doped AuNT with various Pd thicknesses upon in-situ TEM heating up to 1000°C is studied using HR(S)TEM imaging and STEM-EDS. The initial AuNT@Pd core@shell nanostructures show 3D growth of columnar Pd on the AuNT. The height of the 3D structures depends on the amount of Pd salt added during the synthesis. With heating, the 3D structures immediately deform and a homogeneous Au-Pd alloy structure is produced as the temperature reaches 400°C. With further increments in temperature, the nanostructure shows a unique phase separation process between Au and Pd, resulting in nanostructures that are highly stable at RT. The degree of the Pd separation is observed to be in proportion to the amount of initial Pd concentration. With in-situ TEM, we could successfully probe this separation process, which was previously only predicted using theoretical studies. This study helps to understand the phase separation process of two metals, which are considered to be miscible in the entire temperature and composition range and also provides a clear way to understand the effect of the Pd shell thickness for high-temperature application of Au@Pd core@shell nanostructure. The thermal annealing of the AuNT@Pd core@shell nanostructure finally produces a very stable phase-separated Janus NP, which could be useful for different catalytic application purposes.

## 5. Experimental Section

*Chemicals:* Gold chloride trihydrate ($HAuCl_4 \cdot 3H_2O$), sodium hydroxide (NaOH) pellets, Potassium Iodide (KI), L-ascorbic acid (AA), and cetyltrimethylammonium chloride (CTAC, >95%) are purchased from Sigma-Aldrich. The glasswares were cleaned using ultrasonication in ethanol, iso-propyl alcohol (IPA) and de-ionized (D.I) water for 5 minutes, respectively. Followed by drying using $N_2$ gas purging. Millipore water (18.2 MΩ) was used to perform all the synthesis, and all the chemicals were used as received without any further treatment.

*Preparation of Au nanotriangle:* Au nanotriangles (AuNT) were produced by a previously described seedless method adopted by Chen et al.[42] Therefore, 1.6 ml CTAC solution was mixed with 8 ml D.I water in a 15 ml Borosil glass under constant stirring (400 rpm). Next, KI (75 µL, 10 mM), $HAuCl_4, 3H_2O$ (80 µL, 25.5 mM) and 20.5µL NaOH (conc. 0.1 M) were added rapidly in succession with constant stirring (550 rpm), which resulted in a formation of slightly yellowish solution. Then, 80 µL AA solution was injected rapidly into the system, which makes the solution colourless due to the reduction of $Au^{3+}$ to $Au^0$. Finally, 10 µL NaOH (0.1M) was added to the solution rapidly and the stirring was stopped after 2-3 seconds. The colour of the solution became light pink initially and finally dark blue indicating completion of nanotriangle formation. To refine the gold triangle, a 25 wt % CTAC solution was introduced into 10 mL of the prepared nanotriangle solution, achieving a CTAC concentration of 0.15 M. The mixture was left overnight. Following flocculation, the supernatant was eliminated, and the resulting precipitate was isolated and preserved in 5 mL of deionized water. For surfactant removal, the solution underwent dual centrifugation at 7000 rpm for 25 minutes each, with rinses using deionized water and ethanol. Ultimately, the purified solution was stored in 10 mL of deionized water for further use.

*Deposition of controlled Pd layer:* $H_2PdCl_4$ was made by dissolving 44.5 mg $PdCl_2$ with 25 mL HCl solution at RT. Then, for controlled Pd deposition, 0.5 mL AA (conc. 0.1M) was injected into 10 ml AuNT solution under constant stirring, followed by addition of $H_2PdCl_4$. The amount of $H_2PdCl_4$ is varied according to the desired thickness of the Pd layer, which yielded different colours of the solution (AuNT and 13 at%Pd shows deep blue, 36 at% Pd doped AuNT shows purple and 48 at% Pd doped AuNT showed black colour, respectively). The stirring is continued for 10h and, finally, the Pd coated AuNT were obtained by two times centrifugation in DI water (5500 rpm, 10 minutes). The NPs were kept in de-ionized water until further use.

*UV-Visible characterization:* Initially, the NPs were characterized using a JASCO 750 UV-Visible spectrometer in extinction mode. The solution was placed in a quartz cuvette and the measurements were taken for a 300-800 nm wavelength range in extinction mode.

*TEM characterization*: For conventional TEM characterization, the samples were drop-casted on holey C-coated Cu grids. We used a method described by Chen *et al.* to effectively remove the organic contamination (or surfactants) from the AuNT.[79] Two kinds of aberration-corrected microscopes were used for the TEM characterization. Aberration-corrected HRTEM imaging and selected area electron diffraction (SAED) analyses were carried out in an image-corrected TITAN$^3$ (Thermo Fisher Scientific). HRSTEM imaging and STEM- energy-dispersive X-ray spectroscopy (EDS) analyses were carried out in a probe-corrected TITAN Low Base (Thermo Fisher Scientific) equipped with a high-brightness field emission gun (XFEG) and an Oxford Instruments Ultim X-MaxN 100TLE detector for EDS measurements. All the TEM measurements were performed at 300 kV.

*In-situ TEM measurement:* A DENSsolutions Wildfire nanochip based double-tilt heating holder is used for the in-situ heating TEM measurements with commercial microchips that allow heating up to 1300°C. The nanochips have to be coated with a thin layer (<10 nm) of amorphous C at the backside as the heat conductivity of the employed ultra-thin SiN$_x$ membrane was apparently not sufficiently high to guarantee a homogeneous temperature distribution over the membrane area. For in-situ measurements initially a solution of extinction co-efficient 0.5 is obtained for all the NP samples and stored in a 15 ml centrifuge tube. The solution is then diluted 200 times by mixing 10 µL of the nanoparticle solution with D.I. water to finally make a solution of 2ml. The in-situ chips are hydrophobic in nature and to make it hydrophilic, we expose them to plasma cleaning in an LEICA EM ACE 200 plasma cleaner (LEICA Microsystems) and then drop-casted the samples on the chips. The chips are dried overnight, and no further cleaning procedure was done before the in-situ measurements. During the in-situ heating process, the samples were heated to a desired temperature (the transition between the temperature is an instant one) and kept there for 20 minutes to give the NP system sufficient times to undergo the changes. The NPs were continuously tracked and imaged at a particular time interval to get a better idea on the transformation process. The EDS spectra and imaging were obtained after cooling the in-situ holder at RT (beyond measurements of 500°C) as at high T, the X-ray counts can saturate the EDS detector.

**Supporting Information**

Supporting Information is available from the Wiley Online Library or from the author.

**Author Contributions**

The manuscript was written through contributions of all authors. All authors have given approval to the final version of the manuscript.

**Acknowledgements**

These works have been supported by funding from the EU Horizon 2020 programme (MSCA: GA No. 101109165), the Spanish MICIU (PID2019-104739GB-100/AEI/10.13039/501100011033, PID2023-151080NB-I00/AEI/10.13039/501100011033 and CEX2023-001286-S MICIU/AEI /10.13039/501100011033) and by the DGA project E13-23R. The National Centre for Electron Microscopy (ELECMI, Spanish National Singular Scientific and Technological Facility, MICIU) is also acknowledged for provision of access to corrected aberration microscopy facilities through ELECMI Competitive Open Access Protocol

**Conflict of Interest Statement**

The authors declare no competing financial interest.

**Data Availability Statement**

The data that support the findings of this study are available from the corresponding author upon reasonable request.

**Ethical Statement**